\documentclass[floatfix,reprint,prb,superscriptaddress,twocolumn,showpacs]{revtex4}
\usepackage{graphicx}
\usepackage{amsmath}
\usepackage{amsfonts}
\usepackage{amssymb}
\usepackage{amsbsy}
\usepackage{hyperref}

\begin{document}
\title{Helical scattering signatures of strain and electronic textures in YbFe$_2$O$_4$ from three-dimensional reciprocal space imaging}
\date{\today}
\author{Alexander J. Hearmon}
\email{a.hearmon@physics.ox.ac.uk}
\affiliation{Clarendon Laboratory, Department of Physics, Oxford University, Parks Road, Oxford, OX1 3PU, UK}
\affiliation{Diamond Light Source Ltd., Didcot, OX11 0DE, UK}
\author{Dharmalingam Prabhakaran}
\affiliation{Clarendon Laboratory, Department of Physics, Oxford University, Parks Road, Oxford, OX1 3PU, UK}
\author{Harriott Nowell}
\affiliation{Diamond Light Source Ltd., Didcot, OX11 0DE, UK}
\author{Federica Fabrizi}
\affiliation{Clarendon Laboratory, Department of Physics, Oxford University, Parks Road, Oxford, OX1 3PU, UK}
\author{Matthias J. Gutmann}
\affiliation{ISIS Facility, Rutherford Appleton Laboratory, Didcot, OX11 0QX, UK}
\author{Paolo G. Radaelli}
\affiliation{Clarendon Laboratory, Department of Physics, Oxford University, Parks Road, Oxford, OX1 3PU, UK}

\begin{abstract}
The insulating ternary oxide YbFe$_2$O$_4$ displays an unusual frustration-driven incommensurate charge-ordering (CO) transition, linked to possible ferroelectricity. Based on high-resolution synchrotron data, we report a detailed structural model showing that the CO phase is an incommensurate charge-density wave and cannot be ferroelectric, since the electrical dipole moments are also incommensurately modulated. The change between continuous and ``spotty'' helices of scattering at the CO transition is attributed to three-dimensional fluctuations of the direction of the ordering wavevector.
\end{abstract}

\pacs{77.80.-e, 71.45.Lr, 61.05.cp, 75.85.+t}

\maketitle

\section{Introduction}
\begin{figure}[b]
\begin{center}
\includegraphics[scale=0.475]{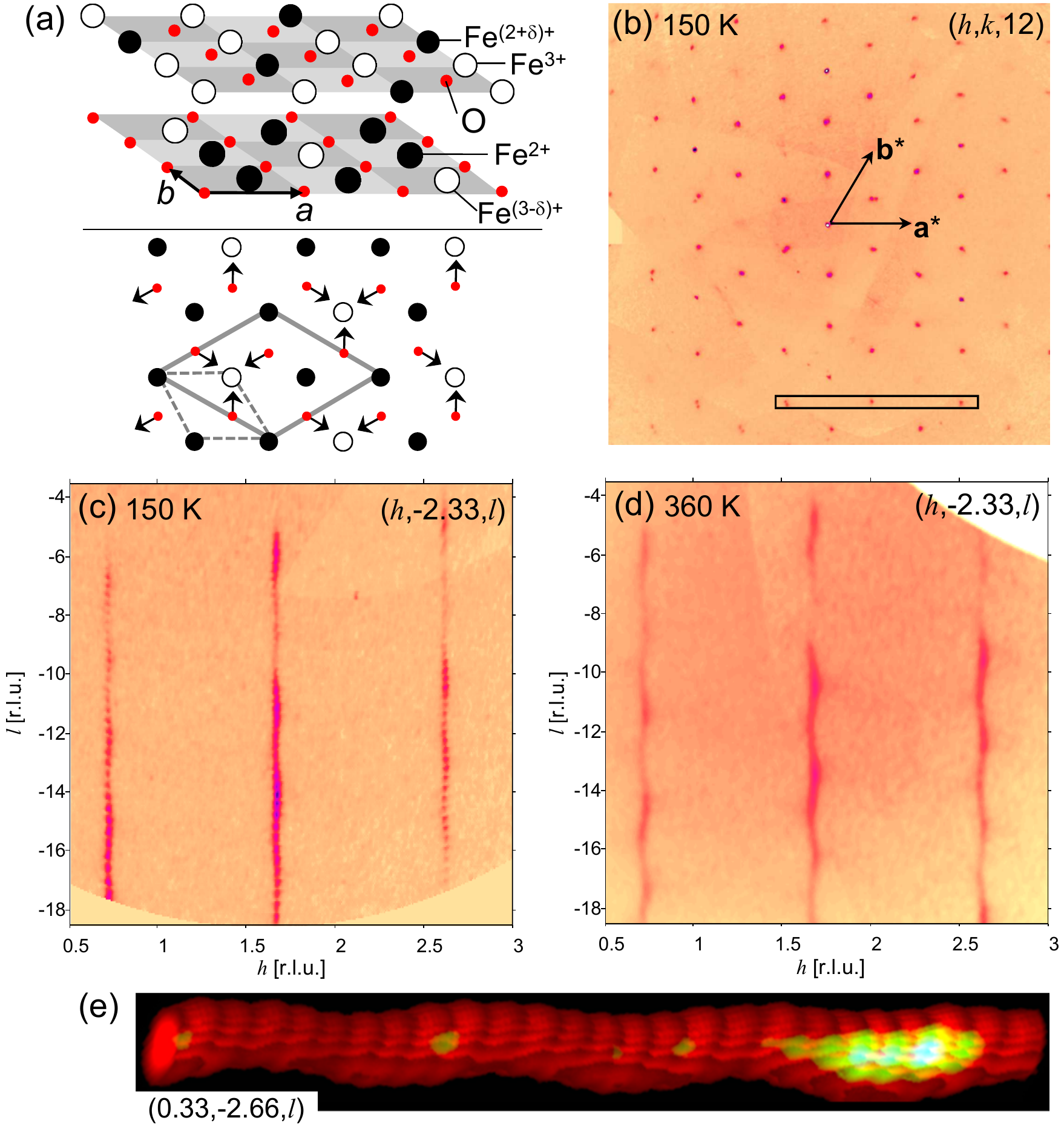}
\caption{(Color online) \textbf{(a)} The simplest ordering model of Fe$^{2+}$ and Fe$^{3+}$ on adjacent FeO layers in $R$Fe$_2$O$_4$ \cite{ikeda2005ferro_32}. Lower panel: oxygen displacement pattern in a single FeO plane. The arrows can also be used to illustrate the conventional $120^\circ$ magnetic structure of a triangular Heisenberg magnet. The CO and crystallographic unit cells are indicated. \textbf{(b)} \& \textbf{(c)} Reciprocal space sections from lab data collected on YbFe$_2$O$_4$ at 150 K, respectively the $(h, k, 12)$ and $(h, -2.33, l)$ layers. Bragg peaks in (b) are subject to the selection rule $-h + k + l = 3n$ and are surrounded by hexagons of satellites. \textbf{(d)} Synchrotron data showing the $(h, -2.33, l)$ layer at 360 K: the streaks observed at low $T$ (b) become more diffuse for $T > T^*$. The satellites shown in (c) and (d) are marked with a box in (b). \textbf{(e)} Three-dimensional iso-intensity plot of the $(0.33, -2.66, l)$ satellite at 150 K (synchrotron data). \label{fig1}}
\end{center}
\end{figure}

The family of systems with general formula $R$Fe$_2$O$_4$ ($R$ = Dy--Lu, and Y) has attracted significant interest, due to the possibility of high-temperature multiferroic behaviour arising from charge ordering \cite{ikeda2005ferro_32,Nagano2007electric_77,Ikeda2005charge_31,angst2008charge_37,ikeda1998ic_73,oka2008charge_35,Xiang2007charge_88,Park2007dynamic_87,Naka2008magnetodielectric_86,subramanian2006giant_82}. Just above room temperature these systems undergo a metal-to-insulator (MI) transition, associated with sharp anomalies in the specific heat \cite{angst2008charge_37}, and on further cooling magnetic ordering sets in at around 250 K \cite{christianson2008three_38,Iida1993magnetization_74,Yoshii2007magnetic_65,Phan2010complex_261}. The MI transition has been interpreted as a peculiar form of charge ordering (CO), where the mixed-valence character of iron (the average Fe valence is 2.5+) competes with the geometrical frustration of the lattice. The crystal structure of $R$Fe$_2$O$_4$ \cite{kato1975kristallstruktur} consists of an alternating stacking of $R\textrm{O}_2$ layers and Fe$_2$O$_2$ bilayers (with three such bilayers per conventional unit cell). The Fe ions form a triangular pattern within the layers, so the system is geometrically frustrated with respect to both the Fe spins and valence \cite{Nagano2007spin_79} and it is impossible to order an equal amount of Fe$^{3+}$ and Fe$^{2+}$ on the same layer. Yamada and coworkers \cite{yamada1997ic_72, Yamada2000charge_40} proposed a possible CO structure for LuFe$_2$O$_4$, which they predicted to be an incommensurate charge-density wave (CDW) and an incommensurate antiferroelectric. Although their work is mainly based on a theoretical model (see also below), they were able to explain the incommensurate positions of the CO satellites which, in reciprocal space sections, `meander' around the lines of commensurability. Ikeda and coworkers later reported that the low-temperature insulating phase of LuFe$_2$O$_4$ is ferroelectric \cite{ikeda2005ferro_32}, and proposed a much simpler \textit{commensurate ferroelectric} CO model in which two inequivalent lattices would be formed within the bilayer: $\textrm{Fe}^{3+} + 2\textrm{Fe}^{(2+\delta)+}$ on one layer and $\textrm{Fe}^{2+} + 2\textrm{Fe}^{(3-\delta)+}$ on the adjacent one (Fig. \ref{fig1}a), creating net dipole moments for each bilayer, which would stack ferroelectrically. More recently, Angst \textit{et al.} \cite{angst2008charge_37} presented an alternative model, based on  LuFe$_2$O$_4$ diffraction data and a commensurate approximation of the charge-ordered superstructure, and proposed that the bilayer stacking is \emph{antiferroelectic} rather than ferroelectric. 

In this paper, we present a set of three-dimensional reciprocal space data on YbFe$_2$O$_4$. We focus on the $R = \textrm{Yb}$ system since we found it to have more stable oxygen content (to which the properties of the system are very sensitive \cite{mori2008effect_71,Kishi1982magnetisation_68}) and to yield better-quality single crystals than $R = \textrm{Lu}$. These data clearly show that the meandering lines of incommensurate satellites are actually helices, as predicted theoretically \cite{Yamada2000charge_40, Harris2010PRBcharge_54}. We also present a real-space model of the fully incommensurate CO structure, which, in spite of its simplicity, is in very good agreement with the experimental data. We show that, below $T^* \approx 320$ K, YbFe$_2$O$_4$ forms a true incommensurate CDW (as predicted by Yamada) associated with an incommensurate rotating strain pattern. Although local electric dipole moments are present, their directions reverse periodically, implying that each bilayer has a zero net moment. Therefore, the previously observed electrical and dielectric anomalies are most likely associated with CDW physics rather than with ferroelectricity. Above $T^*$ the individual superlattice peaks melt into continuous helices of scattering, which nonetheless remain very coherent. This may indicate that low-lying incommensurate states with different wavevectors are becoming progressively populated --- a situation that is not observed in conventional CO transitions.

\section{Experimental methods}
YbFe$_2$O$_4$ single crystals were grown by the optical floating-zone technique. Stoichiometric mixtures of Yb$_2$O$_3$ and Fe$_2$O$_3$ were ground together and calcined in CO:CO$_2$ (4:5) mixed gas flow at 1160$^\circ$C for 48 h with intermediate grinding.  The feed and seed rods (10 mm diameter) were finally sintered at 1175$^\circ$C for 36 h in CO:CO$_2$ (1:2) mixed gas flow. The crystal growth was carried out in CO:CO$_2$ mixed (1:4) atmospheric flow using a IR-heated image furnace (Crystal System Inc.). The feed and seed rods were rotated in opposite directions at 25 rpm during crystal growth at a rate of $\sim$2 mm/h. Preliminary X-ray diffraction data were collected on an Agilent Technologies 'Supernova' diffractometer, using a Mo source (wavelength 0.71073 {\AA}) and cryostream to cool to 150 K. Synchrotron X-ray diffraction experiments were carried out on beamline I19 at the Diamond Light Source, UK. The beamline employs a double bounce monochromator and focussing mirrors to bring the beam down to approximately 170 $\mu$m $\times$ 85 $\mu$m at the sample position. A wavelength of 0.6889 {\AA} was used throughout and data were collected using a Rigaku Saturn 724 CCD detector at temperatures between 150 and 360 K. Reciprocal space sections were generated from the raw CCD images using the programme CrysAlis$^\textrm{Pro}$. Stacks of such sections were converted to a volume data set and analysed using FIJI \cite{FIJIref} to enable visualisation of the scattering intensities in three-dimensional reciprocal space. 

\section{Results}
\begin{figure}[b]
\begin{center}
\includegraphics[scale=0.525]{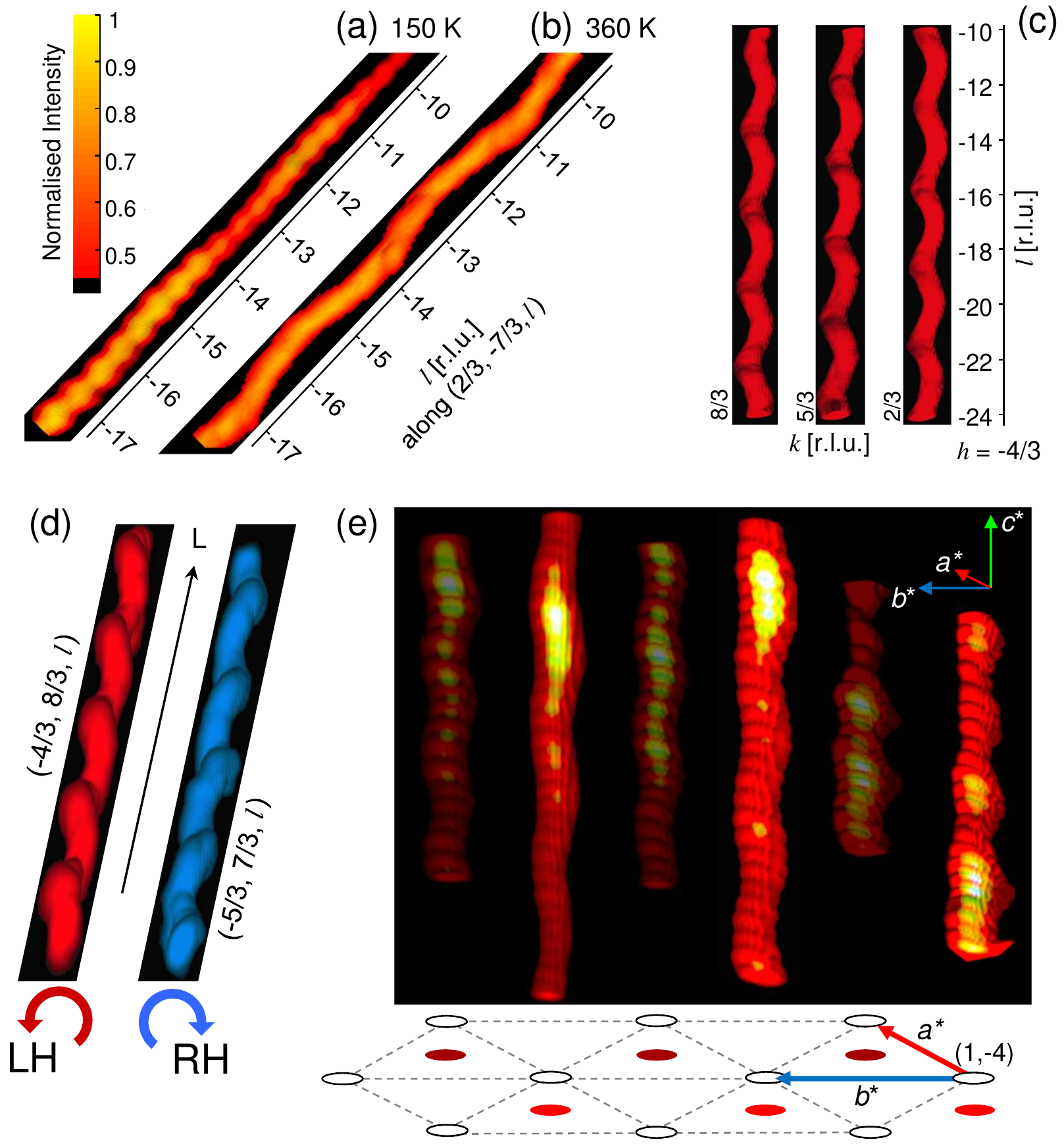}
\caption{(Color online) Three-dimensional reciprocal space iso-intensity plots of the satellite helices. \textbf{(a)} and \textbf{(b)}:  Helices below and above $T^*$. At low $T$ intensity is concentrated onto peaks separated by $\sim 1/3$ $c^*$. \textbf{(c)} Three adjacent helices ($T = 360$ K). The pitch of each helix is 3$c^*$, but the phase varies according to the position in the $(a^*,b^*)$ plane. \textbf{(d)} Two helices at $T = 360$ K, for $24 < l < 10$ r.l.u., illustrating the selection rule for the handedness (denoted RH/LH for right/left). \textbf{(e)} Schematic reciprocal space reconstruction of several satellites at 150 K. The reciprocal lattice is shown below (white circles) with the locations of the satellites (red/dark circles) projected downwards. \label{fig2}}
\end{center}
\end{figure}

Figs. \ref{fig1}b-\ref{fig1}d show reciprocal space sections of X-ray diffraction data collected from both the laboratory source and synchrotron. The $(h,k,12)$ section (data at 150 K are shown in Fig. \ref{fig1}b; the same section at 360 K looks very similar) clearly shows the appearance of satellite spots, which can be indexed approximately on a $\sqrt{3} \times \sqrt{3}$ supercell. A perpendicular section through these satellites (Fig. \ref{fig1}c) shows that the scattering is densely distributed around $(m+0.666, -2.33, l)$ lines, with $m$ an integer. The Bragg peak maxima are not at commensurate positions, but oscillate in $h$ and $k$ around these lines and occur at incommensurate values of $l$, separated by $\sim 1/3$ $c^*$. Above $T^*$ (Fig. \ref{fig1}d), the Bragg peaks `melt' into a continuous meandering line (this intensity is reported to vanish above $\sim 500$ K \cite{Nagano2007spin_79} where CO is no longer present), but remain very coherent perpendicular to the line itself. The data in these sections are very reminiscent of previously published LuFe$_2$O$_4$ data (for example ref. \onlinecite{angst2008charge_37}), although in LuFe$_2$O$_4$ the spacing  along the $l$ direction is $1/2$ $c^*$. However, our high-resolution synchrotron measurements were suitable for full three-dimensional reconstructions (rather than simple sections), affording much richer views of the experimental data. The three-dimensional iso-intensity contours of scattering for one rod (Fig. \ref{fig1}e) and for a larger region of reciprocal space (Fig. \ref{fig2}) evidence that the meandering lines in Figs. \ref{fig1}c and \ref{fig1}d are nothing but sections through helices of scattering in reciprocal space. The radius of the helix, $\rho$, which is of the order $0.025 a^*$ at 150 K (and approximately twice this value at 360 K), indicates that the $\sqrt{3} \times \sqrt{3}$ supercell is only a commensurate approximation to a complex incommensurate CO superstructure. Figs. \ref{fig2}a and \ref{fig2}b show a three-dimensional view of the `spotty helices' below $T^*$ and of the continuous helices above $T^*$. 

These deviations from the commensurate structure arise because the commensurate arrangement of charges proposed by Ikeda et al. \cite{ikeda2005ferro_32} is energetically unstable due to the frustrated topology of the rhombohedrally-stacked bilayers. Rastelli and Tassi \cite{rastelli1986rhombohedral_107} first explored the effects of this frustration in the context of magnetic ordering in solid oxygen, which displays a topologically analogous stacking of magnetic triangular layers, and found that the `conventional' $120^\circ$ magnetic ordering, with propagation vector (1/3, 1/3, 0), is unstable. Instead, if interactions between the planes are taken into account, the system displays a line of degenerate ground states, corresponding to reciprocal space $\mathbf{q}$ vectors lying on the contour of a \textit{helix} that surrounds the (1/3, 1/3, 0) point. In his previously cited analysis \cite{yamada1997ic_72, Yamada2000charge_40}, Yamada \textit{et al.} exploited the topological similarity between the $R$Fe$_2$O$_4$ family and the magnetic systems considered by Rastelli and Tassi, and modelled the `wavy' pattern of diffuse scattering and Bragg peaks observed in LuFe$_2$O$_4$ in terms of reciprocal space `helices'. More recently, Harris and Yildirim \cite{Harris2010PRBcharge_54} constructed a complete mean-field theory of the CO in $R$Fe$_2$O$_4$, and similarly to the magnetic case they concluded that the mean-field ground state (calculated using interactions up to the third neighbour) results in a degenerate 'helix' of $\mathbf{q}$-vectors. A fourth-neighbour term would stabilise a single ground state, which should display Bragg peaks at $h=1/3+\delta_1, k=1/3+\delta_2$ and $l$ either integer or half-integer depending on the sign of the interaction. The mean-field models in refs. \onlinecite{Yamada2000charge_40} and \onlinecite{Harris2010PRBcharge_54} are best compared to our high-temperature data, since their fully ordered phases (with four interactions) have Bragg peaks in different places along the helix. 

\begin{figure}[b]
\begin{center}
\includegraphics[scale=0.5]{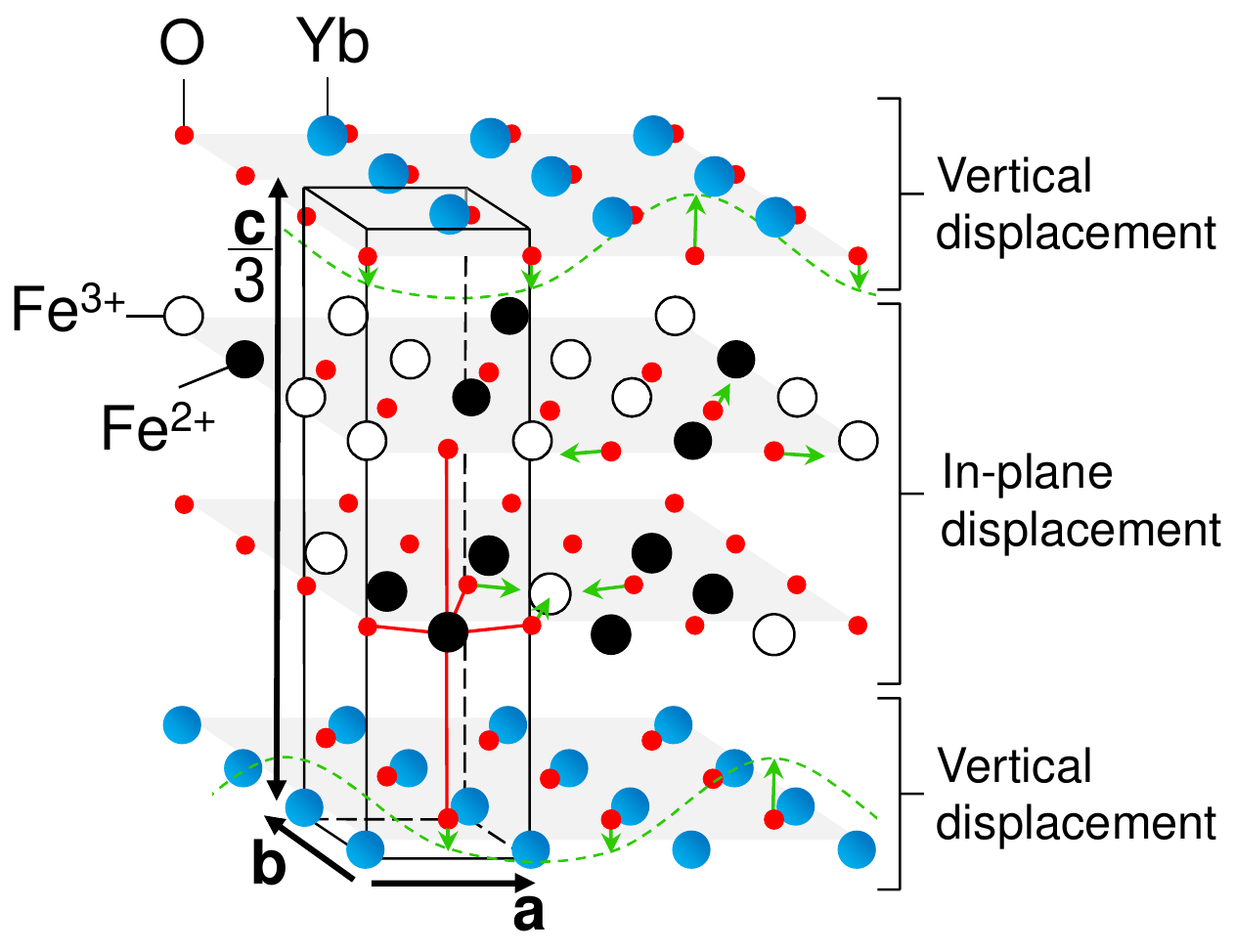}
\caption{(Color online) Schematic picture of the oxygen displacements within the YbFe$_2$O$_4$ structure. Apical oxygens (within the YbO$_2$ layers) are displaced vertically according to a sinusoidal modulation (dashed line), whilst the oxygens in the Fe$_2$O$_2$ layers are displaced in plane (part of the displacement pattern is shown by the arrows). As shown, the modulations follow the commensurate $\sqrt{3} \times \sqrt{3}$ structure. The same modes are propagated with incommensurate $\mathbf{q}$ to generate the final structure.\label{fig3}}
\end{center}
\end{figure}

\begin{figure}[b]
\begin{center}
\includegraphics[scale=0.35]{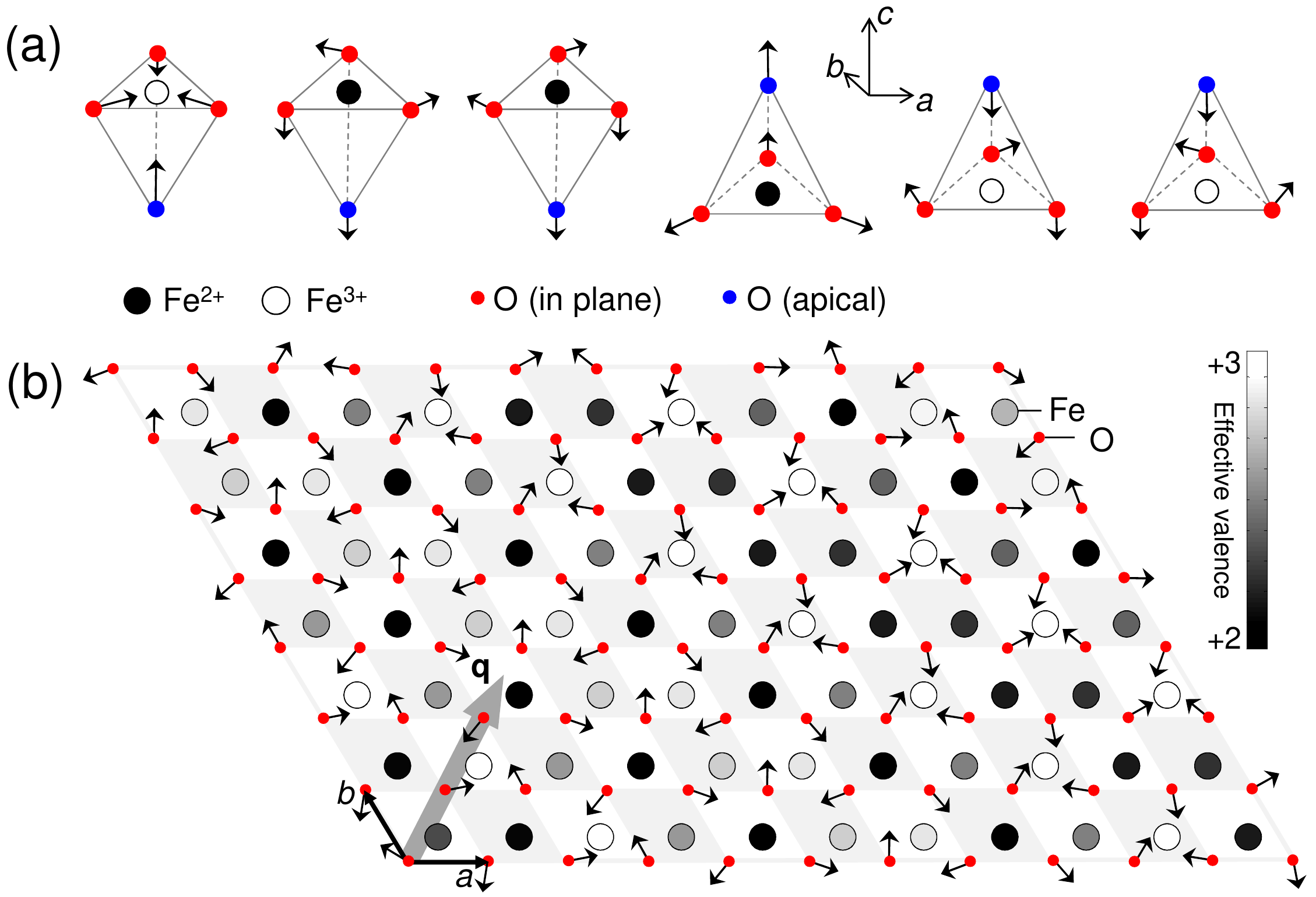}
\caption{(Color online) Real space diagrams of YbFe$_2$O$_4$. \textbf{(a)} The oxygen distortions involved in the six possible pyramids in the $\sqrt{3} \times \sqrt{3}$ structure. \textbf{(b)} Example of an incommensurate oxygen displacement pattern in the FeO planes, made up from a continuous variation of the configurations in (a). The irons are shaded according to their effective valence determined by eqn. (\ref{maxwell}), and the direction of $\mathbf{q} = (1/3 + \delta, 1/3 + \delta)$ is indicated by a bold arrow. \label{fig4}}
\end{center}
\end{figure}

\begin{figure}[b]
\begin{center}
\includegraphics[scale=0.5]{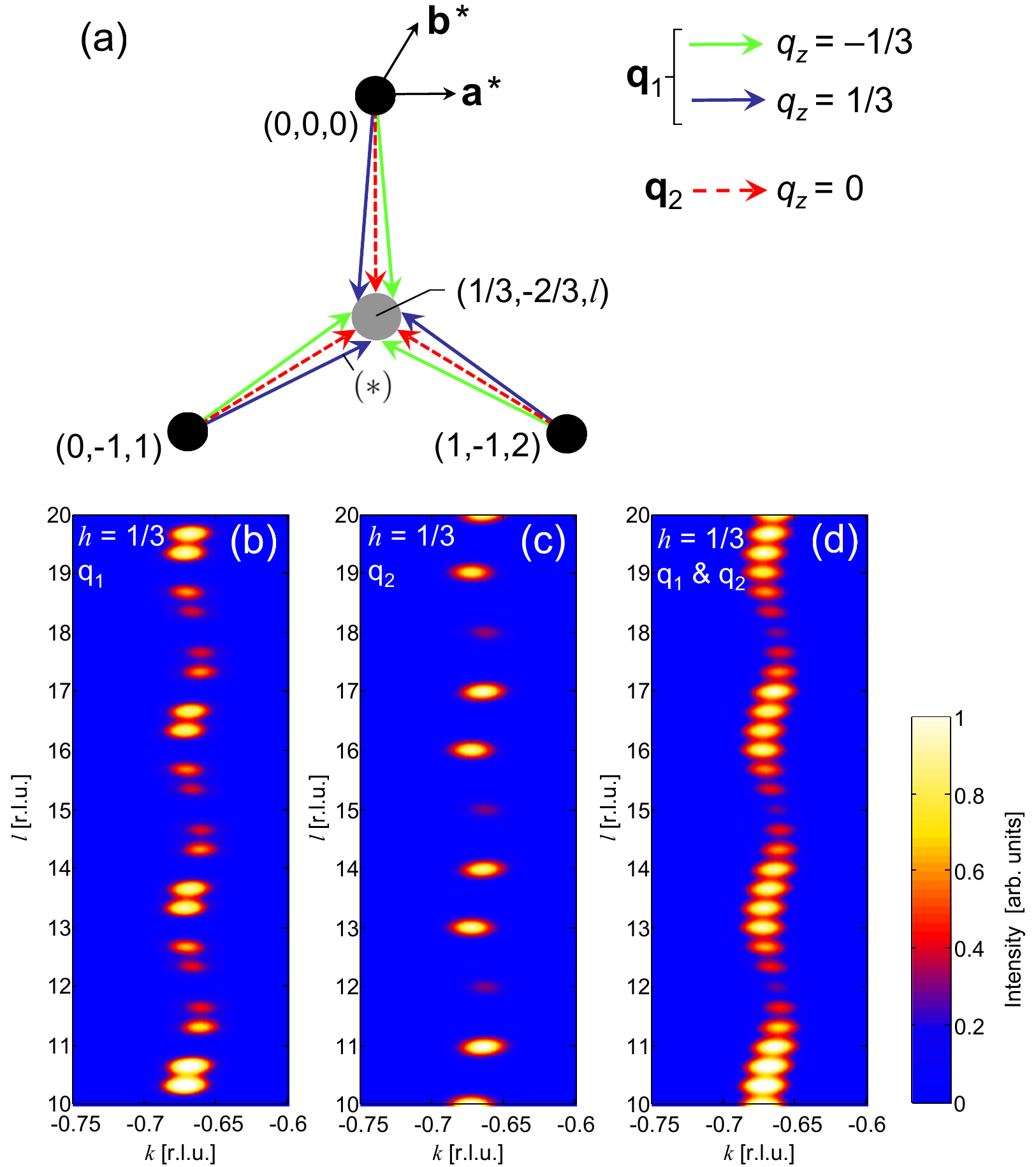}
\caption{(Color online) \textbf{(a)} Reciprocal space sketch indicating contributions from the two propagation vectors, $\mathbf{q}_1$ and $\mathbf{q}_2$, to the helical scattering pattern. The black circles represent Bragg peaks and the grey shows the location of the satellite helix. \textbf{(b)} \& \textbf{(c)} Calculated intensities along the $(1/3, -2/3, l)$ rod from the $\mathbf{q}_1$ and $\mathbf{q}_2$ structures respectively. \textbf{(d)} Overall pattern expected given equal contributions from both structures, which is compared against the data in Fig. \ref{fig6}. \label{fig5}}
\end{center}
\end{figure}

\begin{figure*}
\begin{center}
\includegraphics[scale=0.6]{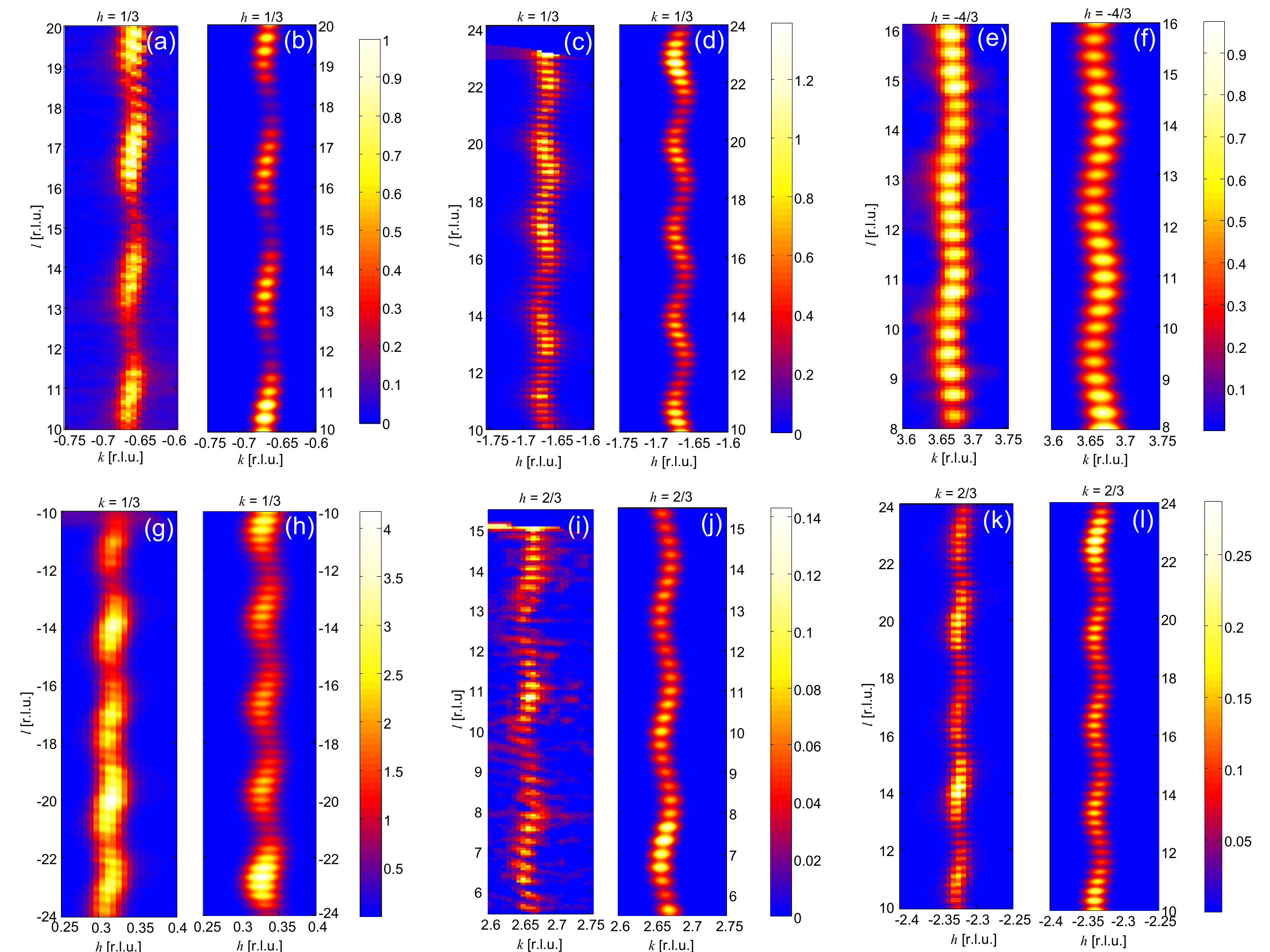}
\caption{(Color online) Reciprocal space sections and simulations at $T = 150$ K. Data (left-hand panels) and calculations (right-hand panels) are shown for six reciprocal space cuts. The calculated intensities are derived from the oxygen displacement model described in the text. The intensities (all of which are scaled to those in (a) and (b)) are indicated by the color bars. \label{fig6}}
\end{center}
\end{figure*}

The topology (handedness, phase, and pitch) of each satellite helix is fixed entirely by the (chiral) holohedral axes that are present along $c^*$ in the rhombohedral reciprocal lattice at $(1/3,1/3)$ type positions. Thus, for a reciprocal lattice of points satisfying $-h + k + l = 3n$ where $n \in \mathbb{Z}$, the chirality of each helix is given by $h+k = (3n \pm 1)/3$ where $+$ $(-)$ corresponds to left- (right-) handed helices (Fig. \ref{fig2}d), and the relative phase, $\Delta \varphi$, of adjacent helices must be $2\pi/3$ (Fig. \ref{fig2}(c)). The pitch of the reciprocal space screw axes additionally fixes the pitch of the satellite helices to 3$c^*$, whilst the \textit{radius} of the helices is an adjustable parameter in the theory, and corresponds to the ratio of the nearest- and next-nearest-neighbour interactions. The 150 K data (Fig. \ref{fig2}a) indicate that at low $T$ unique modes have been stabilised with peaks appearing along the helices with characteristic separation of $\sim 1/3$ $c^*$. Although this is at odds with the theory, the mean-field model could be modified, introducing longer-range interactions, to yield the correct $\mathbf{q}$-vector. The intensities of the peaks also follow the `envelope' of the high $T$ data, indicating that the structural distortions involved are very similar in both phases.

\section{Construction of the incommensurate model}

While our analysis supports Yamada's original proposal, conclusive evidence can only come from a structural model that reproduces not only the Bragg peak positions, but also the observed \textit{intensities}. Although the present data cannot be reliably integrated, preventing a full crystallographic refinement, we present an extremely simple, parameter-free model of the incommensurate CDW that yields very good agreement with the experimental intensity profiles. 

Initially we construct a commensurate displacement (strain) pattern of the oxygen ligands that is consistent with the CO structure proposed by Ikeda \textit{et al.} \cite{ikeda2005ferro_32}. Such patterns, where metal oxygen distances are smaller (larger) for higher (lower) oxidation states respectively, are ubiquitous in CO metal oxides. Ikeda's CO structure can be reproduced by a single \textit{commensurate} wavevector $\mathbf{q} = (1/3, 1/3, 0)$ and two displacements for the in-plane and apical oxygens (along the fixed $\langle 210 \rangle$ and [001] directions, respectively: see Fig. \ref{fig3}). The oxygen displacement pattern in the Fe$_2$O$_2$ layers is completely defined by this rule and by the requirement that the two layers be in phase opposition. This yields six kinds of distorted pyramids, shown in Fig. \ref{fig4}a with the corresponding valence states. A further constraint can be imposed by assuming that the in-plane displacement of each oxygen atom is proportional to the value at its position of a `distortion field' $\mathbf{S}$ obeying the relation
\begin{equation}
\nabla \cdot \mathbf{S} = -\delta \rho
\label{maxwell}\end{equation}
where $\delta\rho$ is the charge (valence) density deviation from the average (this yields essentially equal displacements on the two oxygen sites and a one-parameter model). Eqn. (\ref{maxwell}) is reminiscent of Maxwell's equation for the electric field, but here the displacements are of steric rather than electrostatic origin. The in-plane displacement pattern is the polar analogue of the `conventional' $120^\circ$ magnetic structure: therefore, at a deeper level, eqn. (\ref{maxwell}) defines a relation between scalar and vector ordering on a frustrated lattice -- a concept that underpins, for example, magnetic monopoles in spin ice \cite{castelnovo2008magnetic_121}. A minimal microscopic model of the \emph{incommensurate} phase is then obtained simply by employing the correct incommensurate wavevector, producing, for the in-plane oxygens, the polar analogue of the incommensurate magnetic structure proposed in ref. \onlinecite{rastelli1986rhombohedral_107}. This model yields a continuous variation of the Fe valence on the different sites (with a corresponding variation between the configurations in Fig. \ref{fig4}a). This is shown pictorially in Fig. \ref{fig4}b. The magnitude of the in-plane and out-of-plane oxygen displacements is assumed to be equal for simplicity.

\section{Comparison with scattering data}

The experimental data at 150 K show that the scattering intensity is located at positions along the rod separated by an incommensurate value of $q_z \approx 1/3$. Such an intensity distribution cannot be reproduced with a single wave vector, but requires two symmetry-inequivalent ordering wave vectors (propagation vectors) and their symmetry-equivalent domains as shown in Fig. \ref{fig5}. The first, $\mathbf{q}_1 = (1/3 + \delta_1, 1/3 + \delta_2, 1/3)$, marked with $(*)$ in Fig. \ref{fig5}a, gives rise to 12 equivalent propagation vectors of which six (solid arrows in the figure) contribute to each satellite rod ($\delta_1 > \delta_2$). Considering a structure with only this propagation vector would lead to peaks at $l = n \pm 1/3$ for integer $n$, positioned around the helix. In addition, to obtain scattering at $l = n$ we take a second propagation vector, $\mathbf{q}_2 = (1/3 +  \epsilon, 1/3 +  \epsilon, 0)$ which results in a further three equivalent propagation vectors (dashed arrows in the figure). It seems highly plausible that if the two modes are closely separated in energy then the system may order in such a way that both modes are populated at 150 K.

Scattering intensities due to the two $\mathbf{q}$s were calculated separately as follows. The radius of the helix, $\rho \approx 0.025 a^*$, was used in order to establish commensurate approximations to the incommensurate wave vectors: these were $\mathbf{q}_1 = (0.336, 0.304, 1/3)$ and $\mathbf{q}_2 = (0.320, 0.320, 0)$. For the $\mathbf{q}_1$ structure, a supercell corresponding to this approximate structure was generated and the oxygen atoms displaced according to the model described above. The oxygen displacements are a measure of the CO, which could be estimated, for example, using the bond valence sum method \cite{brown1985bond_146}, provided that both main peaks and satellites are measured. Here, only satellites were measured, and the displacements were nominally taken to be 0.1 {\AA}.

The structure factors were calculated following the standard expression $F(\mathbf{Q}) = \sum f_j \exp(Q_x r_x^j + Q_y r_y^j + Q_z r_z^j)$ where the summation runs over all $j$ atoms in the supercell and the $f_j$ are the atomic scattering factors (which depend on $|\mathbf{Q}|$). The positions of the Yb ions are shifted slightly so as to be at the center of mass of the surrounding oxygens (imposing the periodic boundary conditions associated with the supercell). The resulting intensities (calculated separately for each domain) were then multiplied by an array of two-dimensional Gaussian peaks at the positions given by the propagation vectors to yield the predicted scattering pattern (shown for the $(1/3, -2/3, l)$ satellite helix in Fig. \ref{fig5}b). The width of this Gaussian reflects the coherence lengths of the wave vector domains ($\sim 150$ {\AA} at 150 K). Intensities from the $\mathbf{q}_2$ structure were calculated in an identical way (Fig. \ref{fig5}c). In order to best model the data we assume that an equal population of $\mathbf{q}_1$ and $\mathbf{q}_2$ modes exist in the crystal and therefore simply average over the two intensities to yield the final pattern (Fig. \ref{fig5}d). Given that there is no adjustable parameter other than a single scale factor, the observed scattered intensities along the helix are modelled with surprisingly good agreement over the whole reciprocal space (see Fig. \ref{fig6}), although some of the more subtle features are slightly different, as is to be expected from such a simple model.  

\section{Discussion and conclusion}

Our analysis of high-resolution synchrotron X-ray scattering data collected on YbFe$_2$O$_4$ both above and below  $T^*$ provides new insight into the physics of this class of materials.  Below $T^*$, we can model the helical envelope and intensity distribution of the satellite peaks in reciprocal space using an incommensurate displacement model.   Since the stoichiometry would favour a simple $\sqrt{3} \times \sqrt{3}$ ordering, our observations imply the presence of an intrinsic instability, as proposed in theoretical work \cite{Yamada2000charge_40,Harris2010PRBcharge_54}. Moreover, our model implies a continuous variation of the charge on the Fe ions (between 2$^+$ and 3$^+$, although the amplitude is probably smaller) clearly demonstrating the presence of an incommensurate CDW at 150 K -- to our knowledge the first example of such a structure in a half-doped oxide. Individual layers do not carry any net charge, and the local dipole moments cancel out over the whole bilayer, indicating that neither the ferroelectric nor the stacked antiferroelectric models can be correct. Rather, YbFe$_2$O$_4$ is an incommensurately modulated antiferroelectric, as originally proposed by Yamada. We suggest, therefore, that the electrical data should be critically re-examined. It is well known \cite{GrunerDensityWaves} that the activated de-pinning of CDWs leads to pronounced non-Ohmic behaviour (observed in various organic and inorganic systems \cite{cohen1976nonlinear_124,ong1977anomalous_125,takoshima1980non_126}). The recent transport measurements performed on LuFe$_2$O$_4$ \cite{li2008electric_122} show precisely such an effect which, given the absence of breakdown of CO in an applied electric field \cite{wen2010robust_123}, provides strong evidence for the existence of such a CDW state in this system. The observed pyroelectric currents \cite{ikeda2005ferro_32} could also be an artefact, since metastable states are notoriously ubiquitous in electric-field-driven CDWs \cite{littlewood1982metastability_127}.

The phenomenology of YbFe$_2$O$_4$ \emph{above} $T^*$ is equally intriguing:  we have shown that twisting helices are still present and maintain a high degree of transverse coherence in the $ab$ plane -- at least 70 {\AA} $(20 a)$ -- at all $T$. This diffuse scattering was previously interpreted in terms of two-dimensional ordering \cite{Yamada2000charge_40}; however, the observed diffuse scattering is inconsistent with either intra-layer or intra-bilayer two-dimensional ordering, since these would produce rods, not helices of diffuse scattering. Significant $c$-axis correlations, much greater than the interbilayer distance, must be present in order to produce helices. As previously observed by Yamada, we have shown that the isointensity surfaces appear to follow the contour of the iso-energetic surfaces for modes of given wavevectors, as predicted by the random-phase approximation theory for critical fluctuations \cite{Yamada2000charge_40}. In real space, this would result from a progressive population of modes within the quasi-degenerate helical manifold, starting from those that are closest to the ground state, so that the local coherence is maintained whilst allowing the system to explore states around the minima. We note that this scenario is very different from conventional CO transitions, which are driven by fluctuations of the amplitude of a \emph{single} order parameter.

We acknowledge a series of very useful discussions with Radu Coldea, John Chalker and Pierre Toledano, and assistance with the synchrotron experiment from David Allan, Kirsten Christensen and Robert Atwood.

\end{document}